\documentclass[sigconf]{acmart}

\usepackage{xspace}
\usepackage{soul}
\setstcolor{red}
\AtBeginDocument{%
  \providecommand\BibTeX{{%
    \normalfont B\kern-0.5em{\scshape i\kern-0.25em b}\kern-0.8em\TeX}}}

\copyrightyear{2023}
\acmYear{2023}
\setcopyright{rightsretained}
\acmConference[CHI '23]{Proceedings of the 2023 CHI Conference on Human Factors in Computing Systems}{April 23--28, 2023}{Hamburg, Germany}
\acmBooktitle{Proceedings of the 2023 CHI Conference on Human Factors in Computing Systems (CHI '23), April 23--28, 2023, Hamburg, Germany}\acmDOI{10.1145/3544548.3581347}
\acmISBN{978-1-4503-9421-5/23/04}

\acmConference[CHI '23]{CHI Conference on Human Factors in Computing Systems}{April 23-April 28, 2023}{Hamburg, Germany}
\acmBooktitle{CHI Conference on Human Factors in Computing Systems (CHI '23), April 23-April 28, 2023, Hamburg, Germany} 
\acmPrice{15.00}
\acmISBN{978-1-4503-XXXX-X/18/06}

\newcommand{\rr}[1]{\textcolor{black}{#1}}
\newcommand{\rrr}[1]{\textcolor{black}{#1}}

\def\participants{participants\xspace} %
\acmSubmissionID{2552}

\author{Kimberly Do*\texorpdfstring{$\dagger$}{}}
\affiliation{%
 \institution{Khoury College of Computer Sciences, \\ Northeastern University}
 \streetaddress{360 Huntington Ave}
 \city{Boston}
 \state{Massachusetts}
 \country{USA}
 }

\author{Rock Yuren Pang*}
\thanks{* Listed in alphabetical order. Both authors contributed equally to this research. Corresponding Author: Rock Yuren Pang, ypang2@cs.washington.edu}
\thanks{$\dagger$ Authored during REU at the University of Washington while affilitated with the Georgia Institute of Technology.} 
\affiliation{%
 \institution{Paul G. Allen School of Computer Science, \\ University of Washington}
 \streetaddress{185 E Stevens Way NE}
 \city{Seattle}
 \state{Washington}
 \country{USA}}

\author{Jiachen Jiang}
\affiliation{%
 \institution{Microsoft}
 \streetaddress{}
 \city{Redmond}
 \state{Washington}
 \country{USA}}

\author{Katharina Reinecke}
\affiliation{%
 \institution{Paul G. Allen School of Computer Science, \\ University of Washington}
 \streetaddress{185 E Stevens Way NE}
 \city{Seattle}
 \state{Washington}
 \country{USA}}

\begin{document}

\title[``That's important, but...'': How Computer Science Researchers Anticipate Unintended Consequences \\ of Their Research Innovations]{``That's important, but...'': How Computer Science Researchers Anticipate Unintended Consequences of Their Research Innovations}


\renewcommand{\shortauthors}{Do*, Pang*, Jiang, and Reinecke}

\begin{abstract}
Computer science research has led to many breakthrough innovations but has also been scrutinized for enabling technology that has negative, unintended consequences for society. Given the increasing discussions of ethics in the news and among researchers, we interviewed 20 researchers in various CS sub-disciplines to identify whether and how they consider potential unintended consequences of their research innovations. We show that considering unintended consequences is generally seen as important but rarely practiced. Principal barriers are a lack of formal process and strategy as well as the academic practice that prioritizes fast progress and publications. Drawing on these findings, we discuss approaches to support researchers in routinely considering unintended consequences, from bringing diverse perspectives through community participation to increasing incentives to investigate potential consequences. We intend for our work to pave the way for routine explorations of the societal implications of technological innovations before, during, and after the research process.
\end{abstract}



\maketitle

\section{Introduction}

 From smart glasses that invoke fears of surveillance ~\cite{eveleth_2018} to chatbots that use racist language ~\cite{blog_2016}, our society must increasingly protect itself against the harmful effects of our own technological advancements --- commonly referred to as unanticipated or unintended consequences~\cite{merton1936unanticipated,parvin2020unintended}. While industry is seen as the main offender due to its large user base and broad product impact, computer science research has experienced its own fair share of cases in which innovations have gone awry. For example, AI innovations enabling deepfakes have fostered the spread of disinformation~\cite{kertysova2018artificial}; deep learning models that predict a person's sexual orientation have caused fierce backlash from the LGBTQ community~\cite{wang2018deep}; language models have been shown to exacerbate inequalities~\cite{vieira2021understanding}, propagate social bias~\cite{sap2019risk, koenecke2020racial}, and intentionally produce discriminatory content~\cite{kurenkov_2022}; and innovations in interaction design to improve usability have simultaneously widened disparities between the experiences of the demographic groups included or omitted from the research and development process~\cite{toyama2015geek}. 

 Research in Science and Technology Studies (STS) has documented the hopes and challenges of technological utopianism for decades, but commonly focuses on industry and computer science practitioners, such as software engineers~\cite{kling1991computerization,parker1995all}. Computer scientists in academia may face different and in some ways more complex challenges than practitioners when considering how to anticipate unintended consequences. For example, while their research can result in widely-adopted, non-commercial or commercial products, academic researchers commonly generate ideas and artifacts that are primarily used or extended by other researchers within academia~\cite{Koya2020MeasuringIO}. 
 
 The recent flurry of negative media about the adverse effects of technologies is spurring researchers in various computer science disciplines to more commonly examine ethical implications of their work. Human-Computer Interaction (HCI) researchers have explored the ethics of research and technology in workshops (e.g., ~\cite{bates2019towards,waycott2016ethical}) and various publications (e.g., ~\cite{mackay1995ethics,branham2014co,fiesler2015ethics,mcmillan2013categorised, brown2016five,grimpe2014towards}). Not too long ago, ethics was proposed as one of seven grand challenges for HCI~\cite{stephanidis2019seven}. The SIGCHI research ethics committee has been facilitating open conversations about ethical challenges in our communities through research ethics town halls and panels at CHI ~\cite{frauenberger2017research, munteanu2019sigchi}, CSCW~\cite{bruckman2017cscw, fiesler2021sigchi}, and GROUP~\cite{bjorn2018research}. Similar discussions for more ethical research have been pushed in disciplines such as natural language processing ~\cite{ws-2017-acl, acl}, computer vision~\cite{cvpr2022}, virtual reality ~\cite{Behr2005, Jonathan2021, Ramirez2018}, robotics~\cite{vanderelst2018, ingram2010}, data management~\cite{stoyanovich2022responsible, Abiteboul2019}, and data mining~\cite{hajian2016}. 
 
 Most of the discussions on unintended consequences in computer science research have centered on issues that researchers face during the research process, such as when conducting user studies~\cite{frauenberger2017research}, using online data~\cite{vitak2017ethics,vitak2016beyond}, and crowdsourcing data collection~\cite{Barbosa2019}. What remains unknown is whether and how computer science researchers consider any potential unintended consequences of their innovations on society. How do they incorporate such considerations into their research practice, if at all, and what barriers do they encounter? 

 In this paper, we explored these questions through semi-structured interviews with 20 computer science researchers in various academic positions and sub-disciplines, including Accessibility, Augmented and Virtual Reality (AR/VR), CS Education, Computer Vision (CV), Fabrication, HCI, Machine Learning (ML), Natural Language Processing (NLP), Security, Social Computing, and Robotics. Our findings revealed researchers' current attitudes and practices surrounding the unintended consequences of their research innovations (Section ~\ref{5.1}). Concretely, we observed that researchers recognize the importance of this topic, but do not proactively anticipate potential unintended consequences in practice. In fact, thinking about possible unintended consequences of their research innovations is not an integral part of their research process. To further unpack the reasons for their (in)actions, we identified two main barriers to anticipating unintended consequences: (1) the lack of formal methods and guidelines to anticipate them (Section~\ref{5.2}), and (2) academic practices that promote rapid progress and publications (Section~\ref{5.3}). 

 Overall, \rr{this work presents an in-depth qualitative investigation of applied computer science researchers' current practices and challenges in dealing with unintended consequences.} As a step towards more routinely considering these consequences in computer science research, we discuss directions for future research and provide actionable recommendations to the research community and individual researchers. 

\section{Terminology}
\begin{figure}[h!]
    \centering
    \includegraphics[width=0.4\textwidth]{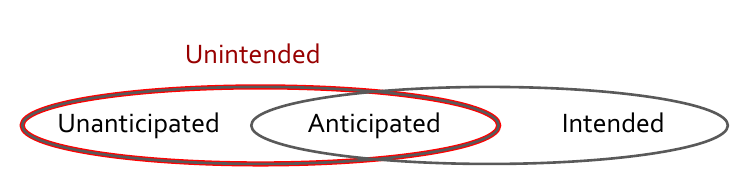}
        \caption{Terminology of anticipated, intended, unintended, and unanticipated consequences. \rr{Unintended consequences can be either unanticipated or anticipated. }}
    \label{fig:term}
    \Description{A Venn diagram with two intersecting circles labeled "unanticipated" and "intended." The intersection is labeled "anticipated". The area including the circle labeled "unanticipated" but excluding the area of the "anticipated" intersection is labeled "unintended."}
\end{figure}

 Already in 1936, Merton~\cite{merton1936unanticipated} coined the term \emph{unanticipated  consequences} to describe unforeseen, desirable or undesirable outcomes of one's action. Today, most researchers refer to unforeseen outcomes (the results of policies, technologies, or other ``purposive social actions''~\cite{merton1936unanticipated}) as \emph{unintended consequences}, though some have suggested that this term conflation has caused a loss of nuance~\cite{huntington1971change, parvin2020unintended}. As we show in ~\autoref{fig:term}, Merton's original term of \emph{unanticipated consequences} suggests that such consequences are always unintended. In contrast, \emph{unintended consequences} can be either unanticipated or anticipated. Parvin and Pollock have therefore argued that this lack of precision in terminology may lead people ``to abdicate responsibility for the perfectly foreseeable consequences of particular decisions''~\cite[p.323]{parvin2020unintended}. Hence, the use of the term \emph{unintended consequences} may reduce accountability: ``Phenomena described as unintended consequences are deemed too difficult, too out of scope, too out of reach, or too messy to have been dealt with at any point in time before they created problems for someone else. The descriptive approach works as a defensive and dismissive strategy''~\cite[p.322]{parvin2020unintended}.

 \label{terminology}
 
 In this paper, we use the term \emph{unintended consequences (UCs)} to purposefully broaden the discussion to include both anticipated and unanticipated, positive or negative unintended side effects of technology on society. Our definition includes consequences that the instigators of an action (i.e., researchers and/or technology innovators) may not have addressed but could have foreseen. While these consequences can be positive or negative in nature (and oftentimes have different effects on a population), our work is inherently oriented towards considering negative UCs more than positive ones. In the remainder of this paper, we use ``technology'' for digital technology, such as hardware devices or software systems. We broadly refer to ``society''  at a regional, national, or international level.

\section{Related Work}
 
 Our work draws upon prior work studying values and ethics in digital technology~\cite{Shilton2018ValuesAE} and is informed by discussions about the effects of digital technology on society in fields such as philosophy~\cite{moor1985computer, johnson1985computer}, STS~\cite{winner2017artifacts, winner2010whale, Klein2002TheSC}, social informatics~\cite{kling1996computerization}, feminism~\cite{Bardzell2010FeministHT,haimson2016constructing} and postcolonial theories~\cite{irani2010postcolonial}. We start by showing how researchers in these fields have long discussed various societal effects of technology before outlining the methods and approaches researchers and practitioners have developed for mitigating unintended consequences. 
 
 \paragraph{Critiques of Technology}
 
 Prior work in STS, and later in HCI, has provided critical analyses of the risks and benefits of technology in society since at least the 1960s~\cite{sveiby2009unintended, Shilton2018ValuesAE}. According to STS scholar Winner, technology ``embodies specific forms of power and authority''~\cite{winner2017artifacts} and technologists should ``pay attention not only to the making of physical instruments and processes [...], but also the production of psychological, social, and political conditions as a part of any significant technical change''~\cite{winner2010whale}. 

 \rr{Work on the risks of technology has been published on a broad range of topics, including the Internet ~\cite{kraut1998internet}, health care information technologies~\cite{harrison2007unintended,ash2007categorizing}, mobile phones~\cite{reyns2013unintended,Moser:2016}, smart technologies~\cite{machidon2018societal}, machine learning~\cite{cabitza2017unintended}, and social media ~\cite{del2016spreading, starbird2019disinformation,starbird2017examining}.}

 The examples above provide a critical lens of the role of technology in society and caution about the unknown and differential effects on societies. Historically, however, most innovation research has focused on desirable and intended consequences~\cite{Rogers1976NewPA, sveiby2009unintended}. In 2009, Sveiby and colleagues suggested that this focus could potentially be due to a ``pro-innovation bias among researchers and vested interests of funding agencies''~\cite{sveiby2009unintended}. Our literature review did not reveal whether this bias has changed in the years since. 

 However, we found many recent calls for more accountability for research innovations~\cite{metcalf2019owning,friedman2019value,Hecht2021ItsTT}. Several prominent computing conferences --- such as the Conference on Neural Information Processing System (NeurIPS ~\cite{NeurIps2021}), Annual Meetings of the Association for Computational Linguistics (ACL) ~\cite{ACL2022}, and the ACM Conference on Intelligent User Interfaces (IUI)~\cite{IUI2022} --- have begun to experiment with ways to encourage or even require researchers to state both the positive and negative potential implications of their work in all paper submissions. \rr{Recent work has made several suggestions for such broader impact statements based on an analysis of these statements in NeurIPS conference proceedings~\cite{nanayakkara2021unpacking, Liu2022ExaminingRA, Ashurst2022AIES}}. After requiring all submissions to contain a section describing the impact of the work, NeurIPS has since transitioned towards a checklist system that offers additional guidance and adaptability ~\cite{NeurIps2021}. 
%

 The HCI community has been raising awareness of UCs of computing research through dedicated publication tracks (e.g., Critical Computing at CHI) and workshops~\cite{conseuqences, unethically}. In particular, a CHI 2021 workshop explored how HCI researchers might think about and report potential negative consequences stemming from their research~\cite{conseuqences}. HCI researchers have also advocated for changes to the peer review process to reduce negative impacts of research innovations, suggesting that reviewers should routinely require that papers and proposals discuss potential adverse effects~\cite{Hecht2021ItsTT}. 
 Given these calls for examining the societal impacts of technology, our work explores whether researchers adopt any methods for anticipating the UCs of their own work. 

\paragraph{Anticipating and Mitigating Unintended Consequences of Technology}
 UCs are often dismissed as unavoidable because anticipating what may happen in the future can be hard~\cite{parvin2020unintended} \rr{and uncertain~\cite{Nanayakkara2020AnticipatoryEA}.} However, HCI researchers have developed ethics-focused design methods to ensure the inclusion of various stakeholders in the design process (for an overview  see~\cite{chivukula2021surveying}). One prominent example is the value-sensitive design (VSD) approach by Friedman, Kahn, and Borning~\cite{friedman2008value}, which can aid in understanding technology, its human value, and its context of use. The process aims to help product teams and researchers identify alternative approaches that better uphold their chosen values while accommodating the same constraints. 
 A number of recent proposals have sought to bridge the gap between theory and implementation by creating toolkits meant for brainstorming about a product's potential societal impacts. For example, the Envisioning Cards~\cite{EnvisioningCards} present the VSD concepts in a clear and modular fashion~\cite{Nathan2008}. In addition, stakeholder tokens also support a VSD stakeholder analysis~\cite{Yoo:2018}. Prior work in HCI has also recommended the use of Tarot Cards of Tech~\cite{Menking:2019} and the Value Cards~\cite{shen2021value} for anticipating potential UCs of specific design choices. 
 
 \rrr{Another approach for considering possible societal impacts is through design fiction~\cite{bleecker2022design, Baumer2018WhatWY}. As a form of speculative design~\cite{Lindley2016PushingTL, Dunne2013SpeculativeED},  design fiction creates a fictional future world to think through sociotechnical issues that have relevance and implications for the present~\cite{wong2017eliciting, lindley2017implications}. This practice has been used to reflect on potential downsides of public data~\cite{fiesler2019ethical}, technology design~\cite{Harrington2022AllTY}, and research prototypes~\cite{soden2019chi4evil}. More recent work developed the design fiction memos method to explore how UX practitioners engage with ethical issues and social impact in their work~\cite{wong2021using}.} 
 
 \rrr{The existing approaches to consider societal implications, however, were often assumed to be effective in practice~\cite{gray2019ethical} and might be difficult to evaluate ~\cite{Baumer2018WhatWY}. While the toolkits often target designers and practitioners as users~\cite{gray2019ethical}, applying them for research projects may pose additional complexities. We extend this line of work by inquiring into whether computer science researchers are aware of and proactively incorporate these tools in their research process. Our work also explores future design implications to support researchers to consider UCs in their research process.
 }
 
\paragraph{Reacting to Unintended Consequences of Technology}
 Not much work has investigated how practitioners and researchers react to UCs in practice. Kling's book on ``Computerization and Controversy'' shows how the power dynamics between programmers and their employers can prevent discussions of potential ethical issues in the products they work on~\cite{kling1996computerization}. As a result, computer science professionals may feel discouraged when reacting to potential or known UCs. 
 \rr{Recently, an interview study showed that the Deepfake open source contributors felt unable to control downstream uses of their software, given the core principle of open source~\cite{widder}.}
%
%
 Researchers have occasionally written public posts in response to public backlash or negative press after deploying a research project~\cite{jiang_2021, openai_2022}, but it is unclear whether they also do so when an incident is less public or when it has only been anticipated (but has not materialized). We fill this gap in prior work by studying whether and how academic computer scientists react if they discover that their work may have UCs. 

\section{Methods}

 In this work, we conducted 20 semi-structured interviews to identify and understand whether and how computer science researchers from diverse sub-disciplines currently approach the potential UCs of their research innovations, what barriers they may encounter, and what design opportunities may exist to support this process. 
 
 \paragraph{Sampling and Participants.} We used purposive sampling to select researchers in computer science who were affiliated with \rr{institutions in North America with very high research activities (R1).} \rr{We focused on North American R1 institutions to reduce potential confounds related to the difference in academic culture and structures, such as funding applications and opportunities, requirements for promotion, and the structure of Ph.D. programs.} Participants were required to work on applied research that has led, or could lead, to systems used by the general public. 
 
 We recruited participants via email after reading their webpages and publications to determine whether their research met our criteria and to ensure diversity in levels of research experience. The email briefly described the research goal of finding ways to support researchers in anticipating UCs. We recruited participants until we reached a sample that satisfied our goal of interviewing computer science researchers \rr{from diverse disciplines and seniority levels} and until the interviews reached saturation. Although we attempted to sample researchers from a variety of applied sub-disciplines, our findings may not encapsulate the thoughts and actions of all computer science researchers. 

\begin{table}[t]
\footnotesize
\centering
\caption{Overview of CS researchers in the study.}
\label{table:reserachers}
\centering
\begin{tabular}{p{0.01\columnwidth}p{0.115\columnwidth}p{0.24\columnwidth}p{0.21\columnwidth}p{0.116\columnwidth}p{0.06\columnwidth}}
    \textbf{\#} & \textbf{Institution} & \textbf{Position}& \textbf{Research Area} & \textbf{Released Public Products} & \textbf{Gender}\\ 
    \toprule
     P1 &  Public & PhD Student & NLP, HCI & No & Male\\
     P2 & Private & Assistant Professor & Social Computing &Yes & Male\\
     P3 & Public & Assistant Professor & ML & No & Male\\
     P4 & Private & Associate Professor & Security, Smartphones, AI & Yes & Male\\
     P5 & Public & PhD Student & NLP & No & Male \\
     P6 & Public & Full Professor & NLP & Yes & Female\\
     P7 & Public & Full Professor & Robotics & Yes & Male\\
     P8 & Public & PhD Student & Computer Vision, ML & No & Male\\
     P9 & Private & PhD Student & AI & No & Male\\
     P10 &Private & PhD Student & AR, VR & No & Female \\
     P11 & Private & Assistant Professor & Brain-Computer Interfaces & No & Female\\
     P12 & Private & Postdoc & Accessibility & Yes & Female \\
     P13 & Private & PhD Student & Fabrication, Sensing & Yes & Male\\
     P14 & Private & Associate Professor &  HCI & Yes & Female \\
     P15 & Public & Assistant Professor & AR, Accessibility & Yes & Male\\
     P16 & Public & PhD Student & CS Education, HCI & No & Female \\
     P17 & Public & PhD Student & CS Education & Yes & Male\\
     P18 & Public & Assistant Professor & AI, Robotics & No & Male\\
     P19 & Public & Assistant Professor & Security & Yes & Male\\
     P20 & Public & PhD Student & NLP & Yes & Female\\
    \bottomrule
\end{tabular}
\end{table}

 Our final sample included 20 computer science researchers (7 female, 13 male) from 10 different academic institutions across North America. All participants had built systems as part of their research, and 9 had released one or more systems as (part of) a public product. Participants held various academic positions in their respective computer science departments (see Table~\ref{table:reserachers}):  10 participants were Ph.D. students or postdocs, and the remaining 10 were assistant, associate, or full professors. Our participants worked in a variety of research areas: 9 participants described their work as being mainly in AI or related areas (e.g. CV, ML, NLP). The remaining participants worked on accessibility, AR/VR, CS education, hardware, social computing, robotics, and security, or a combination of the above. All participants had industry experience, meaning that they have either collaborated, interned, or obtained full-time positions in industry while working toward their research projects.

\paragraph{Interview Protocol.} 

\rr{We prefaced our interviews by loosely defining "unintended consequences" as both desirable and undesirable outcomes of one's research, allowing participants to further elaborate on the term's meaning.} 
 We then divided our interviews into five sections: (1) participant research experience (e.g., research areas, educational and professional background); (2) prior experiences with UCs (e.g., from community norms in their sub-disciplines about considering UCs and/or their own research products have resulted in UCs); (3) understanding whether, when, and how they consider UCs in the research process; (4) understanding barriers to considering UCs in the research process; (5) understanding where researchers perceive opportunities to augment and improve the process to consider UCs. 
 To avoid response bias, we started by explaining that the topic of UCs is relatively new to computer science. In addition to their own experience, we asked questions about habits and norms in researchers' labs and communities to better understand the context for their opinions and avoid participants feeling accused or put on the spot. We used additional questions to probe three topics that came up repeatedly: understanding whether researchers actively anticipated UCs, understanding what may hinder them from doing so, and understanding the design opportunities to support them in anticipating UCs. See Supplementary Materials for the complete list of interview questions. 
 
 Nineteen interviews were conducted remotely over Zoom, and one was conducted in person. All participants consented to being audio-recorded. Each interview was between 30-45 minutes long. We made a financial donation of \$15 per participant to a COVID-19 relief fund to compensate each participant for their time. The study was determined as exempt by our Institutional Review Board (IRB). 

\paragraph{Analysis.} 
Our research team used an inductive thematic analysis process~\cite{braun2006using} where two researchers individually reviewed and conducted open coding on two interviews. Next, three researchers met to create and discuss the first draft of the codebook. Two members of the research team then independently coded several more interviews and refined or added them to the codebook, which was discussed with the full research team. Once consensus was reached, all interviews were re-coded using the revised codebook. 

The final codebook contained 12 top-level codes relevant to how researchers have thought about, experienced, and responded to UCs; it also included codes related to attitudes towards UCs and support researchers needed to think about UCs. Finally, three researchers used affinity diagramming to develop themes based on our codes. 
\rr{Although we discussed both positive and negative UCs in our interviews, the nature of our research led us to focus on participants' reports of negative UCs.} We slightly edited some of the quotes in Section \ref{Results} for readability. 

\paragraph{Positionality.}

 We acknowledge that our academic and professional backgrounds shape our perspectives on this topic. One author teaches computer ethics at an R1 institution. Collectively, we are US-based researchers at two R1 universities and a large US-based multinational corporation. Our academic backgrounds are in Computer Science, primarily as HCI researchers. 
\section{Results}
\label{Results}

Our analysis surfaced three high-level themes. First, we describe current attitudes and practices surrounding the anticipation of UCs (Section \ref{5.1}). We then discuss how the lack of a formal method inhibits the anticipation and reaction to UCs by researchers (Section \ref{5.2}). Finally, we show how academic practices strain efforts to anticipate UCs (Section \ref{5.3}). Participants are identified with a "P". For a small number of sensitive quotes, the specific research products are omitted to provide an additional layer of anonymity.

 \begin{figure*}[ht]
    \centering 
    \vskip 0.3cm
    \includegraphics[width=\textwidth]{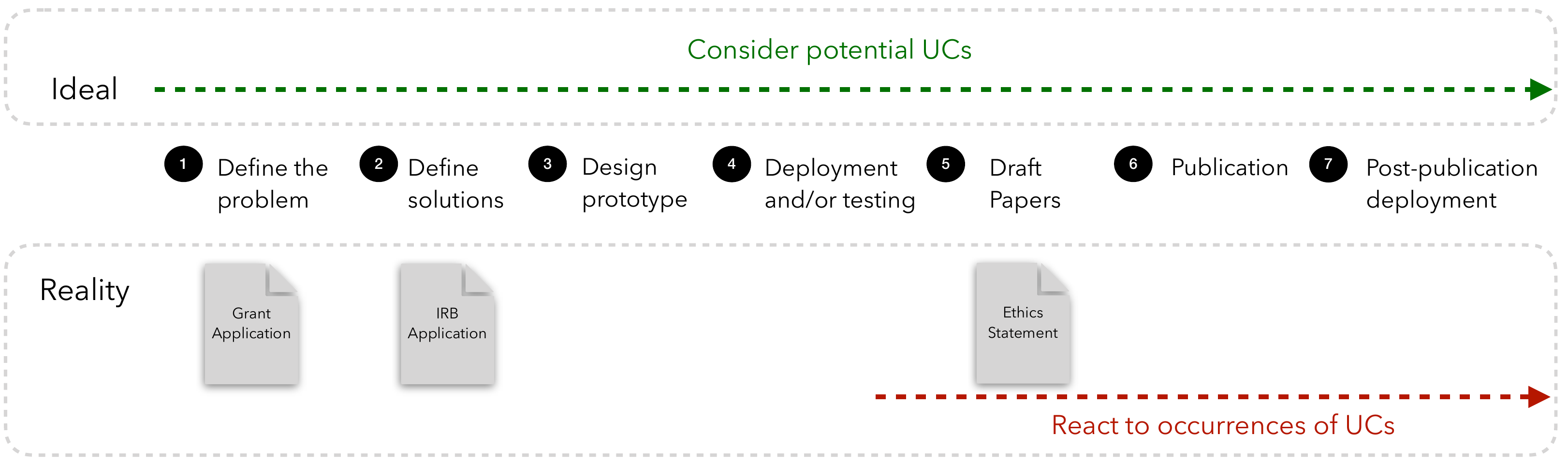}
    \caption{Stages in the research process during which UCs should be considered based on our interview results. Our findings suggest that considering potential UCs should ideally occur throughout the research process (i.e., the dashed green arrow), but that it is currently only done in reaction (i.e., the dashed red arrow) when writing grant applications, IRB applications, and ethics statements, if at all.}
    \label{fig:overview}
    \Description{A timeline that lists the 7 stages of the research process: 1) Define the problem, 2) Define solutions, 3) Design prototype, 4) Deployment and/or testing, 5) Draft papers, 6) Publication, and 7) Post-publication deployment. The ideal time to consider potential UCs is depicted with a dashed arrow that extends through all seven stages of the research process, starting from "Define the problem" and ending at "Post-publication deployment." The reality of considering UCs is depicted with a grant application at the first stage labeled as "Define the problem", an IRB application at the second stage labeled as "Define solutions", an ethics statement at the fifth stage labeled as "Draft Papers", and a dashed line labeled as "React to occurrences of UCs" extending from the fourth stage labeled as "Deployment and/or testing" until the end of the research process.}
\end{figure*}

\subsection{Current Attitudes and Practices Surrounding UCs}
\label{5.1}

In our interviews, 18 out of 20 participants explicitly mentioned that they had at least one prior experience where their research innovations had UCs after deployment or testing. All 20 participants unanimously emphasized the importance and responsibility of researchers to consider UCs throughout the research processes. To illustrate when our participants suggested ideally considering UCs, we present an overview of the research process in~\autoref{fig:overview}. It shows that our participants indicated that UCs should ideally be considered throughout the research process, ranging from problem definition to publication and public deployment. In reality, we observed participants only anticipate UCs when they are required to do so, such as when writing  broader impacts statements for grants, IRB applications, or ethics statements for conferences. We will provide more details describing how participants told us how they use these artifacts for reflecting on UCs in later sections (Section \ref{5.1} and Section \ref{5.2}). 

Despite the ideal circumstances for considering UCs, our analysis showed that none of our participants proactively consider them in the research process. Hence, \textbf{while participants see the importance of anticipating UCs, they rarely take actions to do so}. In fact, 10 participants self-described that they ``do not spend enough time thinking about UCs,'' and 4 participants explicitly mentioned that they ``do not spend any time thinking about UCs.''

Despite not proactively anticipating UCs, our interviews revealed that \textbf{potential UCs are occasionally discussed through informal, serendipitous conversations, but there is no formal process}. Through our interviews, we found that researchers were sometimes made aware of potential UCs through informal conversations with their colleagues and collaborators. They recalled anticipating UCs during lab presentations or meetings, though infrequently. For example, P6 mentioned that collaborators sometimes share concerns, including potential UCs, on a case-by-case basis during project meetings. When asked when they think of UCs in their research, P6 remarked that they do so \emph{``on and off. I mean, the thought is generally there [...] because it's not like the research projects change so often.''} While giving informal research presentations in their lab, P1 welcomed feedback on potential UCs, but bewailed that \emph{``just talking about a research project doesn't necessarily provide an invitation for talking about UCs.''} 
Many researchers shared similar experiences of learning about potential UCs through informal conversations or presentations, but also receiving limited feedback (P1, P4, P6, P7, P9, P10, P12, P16-17). For example, P16 shared her experiences on how \emph{``[Talking about UCs] is kind of something that just happens, during meetings, when if somebody just has a question to discuss, `Oh, I don't know if that's a great idea because of this and that.' ''} However, P16 also expressed that such an approach is \emph{``not like a formal [discussion to] make sure that we're going through every aspect of this tool and making sure it's not going to have these negative consequences.''} Later P16 stated that the informal, or ``accidental,'' conversations on UCs are not an ideal solution. 

In addition to these informal conversations, \textbf{participants commonly react to UCs only after creating tangible research artifacts which can range from deployment and/or testing to post-publication.} Of the 18 participants that shared specific experiences, 13 specifically recounted how experiencing UCs impacted later research decisions, such as making adjustments to a current research project (P5, P7, P17), identifying a new research direction (P12, P14, P19), terminating a research project or idea (P10-11, P13), or consulting expert assistance (P2, P4, P6, P11, P18). For example, P2 shared their experience when testing a content-sharing application in a user evaluation. P2 realized that some audiences found \emph{``some content might not be appropriate for them.''} Therefore, they spent more time on content moderation than originally planned. Similarly, during post-publication, dissemination of results through methods such as social media may also heighten awareness that research innovations can have undesired societal impacts on researchers. For example, P10 told us:

\begin{quote}
    \emph{``People put that [my] research project on Reddit. And there was like, a whole bunch of comments on Reddit on my research video [...] Like, there was something that I didn’t think about [...] And it was kind of something that actually led me to change research topics.''} 
\end{quote}

P20 described their experience with a prototype system that received over 1 million adversarial examples from public auditing. They spent nights debugging the system, replying to users' critiques on Twitter, and issued a public disclaimer in the end. P20 recounted the outcome of their experience: 

\begin{quote}
    \emph{``When we initially released the archive paper [...], we didn't expect a lot of people would just play with a demo, because that's usually how these research prototypes are. [B]ut when we released the demo, suddenly, a lot of people played with it. So yeah, we never thought people would have had such an explosive discussion of this.''} 
\end{quote}

 As a result of reacting to UCs, participants became aware of the impact of their research products on users or society. While our participants took different actions to react to UCs, ranging from issuing public disclaimers (P6, P20), to debugging the system (P1-3, P5, P7-8, P14-16, P20), and shifting research direction (P2, P10), our interviews did not capture a standard procedure for anticipating UCs, such as a dedicated time and research procedure. Nonetheless, some participants regretted considering UCs only after a specific incident had occurred. As P5 put, \emph{``A lot of people do treat [UCs] as an afterthought. And it’s kind of unfortunate.''}

 We also found that many \textbf{participants unintentionally deflect responsibility for considering UCs, believing that their research is unlikely to cause enough harm to warrant serious consideration of UCs}. Under the following subtheme, we describe researchers’ acts of and justifications for this unintentional deflection.

 Although all participants recognized the significance of anticipating UCs, some questioned whether researchers were personally responsible for anticipating the UCs of their research. While none of our participants stated that they ignore UCs altogether, several of them described that they are familiar with colleagues in the research community who disregard their work's societal implications (P6-8). As P8 put it, \emph{``[Some colleagues might think] scientists are not responsible for broader impacts [...] They do research and then let policymakers or someone else think [about them].''} 
 Similarly, P6 asserted that \emph{``there are some machine learning people who think [...] I am going to focus on algorithms and not worry about social issues.''} 

 In contrast, some participants felt that researchers are responsible for the UCs of their work because they are the creators of new technologies. For example, P12 explained that:

\begin{quote}
    \emph{``Researchers and anyone who is bringing these ideas to light, I think they need to be responsible for what they're saying, or at least be able to say whether they have discussed or kind of attempted to figure out what kind of impacts or implications are for whatever information they're presenting to the world.''}
\end{quote}

 Despite the perceived responsibility of some participants, others worry that they lack the influence or foresight to engage with broader societal consequences. For example, P5 described working on a computer vision project that ultimately made it \emph{``easier for the NSA [National Security Agency] to spy on people,''} but they neither anticipated nor reacted to this specific incident. They reconciled themselves with working on it because \emph{``that is kind of why they were funding us.''} Similarly, P20 shared that it is \emph{``extremely difficult to predict what can go wrong without seeing how [the research product] works in the real world, as an NLP researcher.''} Moreover, some participants felt that how their research is used by others is beyond their own responsibility. To them, UCs were unfortunate, yet inevitable repercussions of research, but not necessarily a responsibility to consider or feasibly address.

 We also found that the \textbf{Institutional Review Board (IRB) applications are mistakenly relied upon to alert researchers about the potential for UCs}. When asked about how they consider UCs in their research, some participants shared that they rely on the IRB application (P3, P8, P10-11, P13). P13 thought that \emph{``the [IRB] committee does a good job and eliminates most of the foreseeable negative effects [of their submitted research projects].''} Likewise, P16 expressed similar confidence in the IRB application: \emph{``The whole point of the IRB is that we’re doing things ethically, right? [...] So yeah, I think [considering unintended consequences] falls under the same jurisdiction of [the IRB].''}

 Others were well aware that the IRB process is not designed to anticipate UCs. For example, P10 noted that although the IRB considers \emph{``[unintended] consequences in terms of individual participants, it does not specifically consider general consequences and consequently leaves the responsibility of considering unintended consequences to individual [researchers] to do it}.'' 
 P15 explained that the IRB application might not properly address non-human subjects research and stated that \emph{``many AI projects where you don’t interact with participants''} are not even considered by the IRB. Although the IRB application considers participant ethics, it is not designed to anticipate UCs on society beyond those related to participants. The observation made intuitive sense as the Common Rule, which governs the IRBs in the U.S., specifically disallows review of UCs to human society~\cite{commonrule}.

 As noted above, all participants recognized the importance of considering UCs. Nevertheless, we observed that \textbf{\participants underestimated the potential UCs of their own research because the primary goal of most research is to produce societal benefit}. In general, researchers equated good intentions with producing less social harm. For example, P4 explained that \emph{``there is a pretty big distinction between [my] research [...], which tends to be very specifically targeted towards what hopefully will be societal good,''} and \emph{``technologies that have caused more problems,''} specifying that their intent to create technologies with social benefit lessens the need to scrutinize potential UCs. P16 explicitly stated that \emph{``researchers like us aren't trying to create something that could be harmful.''}
Similarly, P6 noted that:

\begin{quote}
\emph{``By design, my research is really geared towards mitigating the problem that's out there, so it doesn't make sense for me to worry about the negative consequences […] or think about the ethical concerns as much. Because it doesn't seem to apply.''}
\end{quote}

\rr{Although initially provided with a clarifying definition of UCs, multiple participants assumed that the positive, intended consequences of their research would reduce or eliminate its potential negative consequences, leading them to neglect considering UCs.}

Additionally, other participants felt that there was no need for them to consider potential UCs because \rr{they perceived their research as unlikely to be misused}. For example, P5 indicated that they are not as concerned with large-scale UCs because their work is mostly hidden from the public. As an NLP researcher, they explained that \emph{``if anyone wanted to, like really generate fake news, at scale, they probably wouldn't even have used our model.''}
Incidentally, many \participants across multiple sub-disciplines shared this sentiment, but none could explain at what stage their research would require investigating potential UCs. Nearly all researchers explained that because their research created developmental technologies that might lead to \emph{``future work''} [P20], they felt that any potential misuse of their research would not generate sufficient harm for them to need to consider UCs earlier on. In the words of P7, \emph{``because there are no consequences in academia, researchers have to get really worked up about [considering UCs].”}

In summary, we found that our participants perceive considering UCs to be important, but they rarely take proactive steps to address this issue. Instead, they react to UCs when they occur. As P19 said, ``I still think \emph{that's important, but} I just think it's really hard.'' Perhaps due to this attitude, many researchers unintentionally deflected responsibility for four key reasons: reliance on others to consider UCs, dependence on the IRB application to anticipate UCs, belief that research motives equated with reduced social harm, and doubt about the potential social impact of their work.

\subsection{Researchers Lack Formal Methods and Guidelines for Anticipating UCs}
\label{5.2}

Although participants understood the significance of anticipating and responding to UCs in their research, they generally felt unsure of how to approach this issue. Many participants reported that a lack of understanding and experience of UCs reduced their ability to anticipate and/or react to UCs. In this section, we delve into a list of barriers we identified from our interviews: a lack of systematic guidelines for anticipating UCs, a lack of experience and knowledge in UCs which hindered researchers' ability to anticipate and react to UCs, and a lack of opportunities to work with collaborators from diverse backgrounds and skill sets which also hindered efforts to anticipate UCs. 

Many researchers lamented \textbf{a lack of systematic guidelines for brainstorming about UCs and for knowing how to anticipate UCs}. P10 expressed that many conferences do not provide sufficient support --- such as \emph{``predefined infrastructure or scaffolding approach for thinking about them''}, despite requirements for broader impacts or ethics statements. As a result, researchers can feel that \emph{``it's really on the individuals to do [anticipate UCs].''} Compared to existing specific AI-related checklist ~\cite{datasheet, madaio2020co, micrsoft}, a guideline can inform \emph{``at what point you know if we could potentially have some very negative unintended consequences,'' ``who should we bring in,'' ``what kind of outside expert would be the most appropriate for this.'' [P3]} 

Many participants yearned for guidelines to assist with anticipating UCs. P16 noted how a checklist for anticipating UCs could help researchers confirm \emph{``that they thought about [UCs] thoroughly."} In addition, P8, who expected to release an open-source CV demo, expressed a \emph{``need to develop some type of guidelines and policy for how much evaluation is sufficient [before deploying public technologies].''}

Several participants also suggested the benefit of showcasing past UCs that others have anticipated or reacted. For example, P12 acknowledged how they had learned to anticipate UCs based on previous experiences, so \emph{``a resource of common problems [...] of the past might be helpful, so I don't make any of the mistakes that people have already made.''} P10 suggested several potential resources for anticipating UCs including \emph{``some model examples of papers that have done a good job, researchers or research groups that have done a good job, [...] guidelines for how to approach [UCs] to begin with, [...] and thought experiments in thinking about how to how to come up with [UCs].''} Similarly, P19 emphasized the impact of highlighting \emph{``very impactful paper[s] that include ethical statements and win an ACM award''} as a way to not only provide guidance on how to structure writing about UCs, but also serve as \emph{``exemplars of papers''} that demonstrate the importance of anticipating UCs to the academic community.

Similarly, \textbf{the broader impact or ethics statements --- in grant applications and paper submissions --- are insufficient in adequately accounting for UCs.} In grant applications, for example, P12 mentioned researchers might inaccurately portray the impact of their research, which \emph{``is typically framed in a positive light rather than a negative light,''} in an effort to receive funding. The broader ethics statement that some conferences require were also discussed multiple times in our interviews. Although our participants viewed ethics and impact statements as the first step to anticipating UCs, they considered them as “superficial” solutions. P14 explained that, to many researchers, broader impact statements felt more like formalities that \emph{``researchers [have] to do to get publications, because that's what helps them in their careers.''} The broader ethics statement may make researchers only \emph{``think about unintended consequences or these other societal issues when they are writing''}, rather than \emph{``when they are designing studies.''} [P17] Additionally, we note that \participants also feel conflicted about the effectiveness of these statements at guiding researchers to thoughtfully anticipate UCs throughout their research process. P12 worried that simply reporting UCs, rather than acting on them, does not sufficiently protect against the implications of negatively-impacting research. P8 added the limitation from a reviewer's perspective:

 \begin{quote}
 \emph{``[A broader impacts statement] doesn't really do anything, this is just like, a bandage on like a deeper wound of really investigating [...] whether or not a project should be pursued or not [...] Most reviewers when they read a paper to decide to accept to a conference or journal, won't really seriously considered the broader impacts when deciding to accept or reject the paper.''}
 \end{quote}

Moreover, many participants added that \textbf{a general lack of experience and knowledge in UCs hindered their ability to anticipate and react to UCs in their research}. Our interviews indicated that new faculty and junior researchers (e.g., graduate students, postdocs) were most likely to feel unsupported and ill-equipped when anticipating UCs. In particular, several participants shared their experiences with anticipating and addressing UCs as junior researchers. For example, P12 reflected on how, after experiencing a UC in their career, they realized that they did not have the research experience and foresight to anticipate any negative UCs, stating that they ``\emph{[were] overly optimistic [...] about what the realistic [consequences were] going to be [and] sort of unaware of the [potential unintended consequences].''} Other junior researchers also recognized how their limited research experiences prevent them from anticipating UCs; therefore, to anticipate UCs, these junior researchers depend on their advisors to account for potential UCs (P1, P9-10, P13, P17). For example, P13 recounted an interaction with their advisor at the beginning of their Ph.D. where their \emph{``PI basically turned [a research idea] down because it [could] have some negative use cases.''} Some junior researchers shared that they lack opportunities to consider UCs because of their mainly independent work. P1, a Ph.D. student, noted that their lack of collaborators led them to consult \emph{``mostly [the] advisors in [their] labs''} to receive feedback about UCs. Additionally, P17 also acknowledged that ``\emph{maybe others have seen multiple things, or you get more perspectives or more voices. But that's generally not the case for an average Ph.D. student, it's usually a one-to-one relationship with an advisor when they're starting [a Ph.D.]''} 

In particular, junior researchers’ limited research experiences and mostly independent research work often meant that they were unable to properly account for UCs without the assistance of their advisors. Therefore, junior researchers might assume that more senior faculty would have already anticipated negative consequences. For example, P17, a Ph.D. student, explained that his advisor played a large role in screening his proposals for any potential UCs that they were simply unaware of. They explained that:

\begin{quote}
\emph{``I think [advisors] have realized [what the possible unintended consequences are] already [because] they went through the process, through writing what they have seen in the community [...] And if you spend enough time in the community, you get to know those norms implicitly.''
    }
\end{quote}

Further, several participants noted how new faculty also face difficulties when anticipating UCs. P18, a new faculty member, shared their thoughts on confronting UCs: \emph{
``I feel like there's a responsibility there, but I also feel so incredibly ill-equipped to do anything useful about it.''} P19, another junior faculty member, also commented on the knowledge-based limitations that new faculty face in anticipating UCs and questioned their ability to consider \emph{``the ethical reasoning about [a research project] if I'm not personally very educated on it.''} 

\rr{Ironically, several faculty members noted that most students now enroll in ethics courses, and they therefore believed that students are sufficiently prepared to confront ethical problems in their research [P4, P6, P16]. However, our observation indicated that students and even junior faculty members may feel unable and unempowered to approach UCs without the support of their advisors and peers. We recognize how throughout the research pipeline, many researchers — including graduate students and faculty members — face knowledge barriers when anticipating UCs. Ultimately, these findings illustrate the lack of knowledge-based resources that \textit{all} researchers face throughout the research pipeline. }


Exacerbating these knowledge barriers, \textbf{sharing and learning from prior experiences in computer science is largely unsupported, both individually and at a research community level}. Participants stated that reporting UCs during or after publication made it difficult for other researchers to gain awareness of potential pitfalls others have encountered. For example, P11 shared their process for anticipating UCs and explained how they reviewed \emph{``other papers that do similar work to see how they've done if they documented anything, [but] a lot of times they don't.''} Similarly, P19 suggested \emph{``having checkpoints where you share sort of interim progress that you make with the community''} so that other researchers can identify and anticipate when potential UCs could occur during the research process. Notably, P14 and P16, who realized the UCs of their work only after publication and deployment, reported that they could not revise their papers without either retracting the work altogether or adding significantly more contributions to the original work. As P16 explained:

\begin{quote}
\emph{``And now [that] we're seeing that there are some unintended consequences, there should be a way for me as the researcher, so that I have the responsibility to go over to that, wherever it's published, and comment on it and be like, here's an update or whatever, so that the community knows.''}
\end{quote}

To these researchers, reporting UCs was burdensome and potentially harmful to their careers, but they both felt willing and obligated to report these updates. Nonetheless, due to significant structural barriers, none of these researchers were able to share these contributions.

Lastly, many researchers suggested that \textbf{a lack of opportunities to work with collaborators from diverse backgrounds and skill sets also hindered efforts to anticipate UCs}. In general, many researchers believed that more opportunities to work with collaborators from diverse backgrounds and skill sets could better support anticipating UCs by enabling the exchange of different viewpoints (P1-2, P8, P11-12, P14-15). For example, P12 asserted the importance of \emph{``be[ing] able to get feedback from people with different life experiences and expertise.''} \rr{In contrast, several researchers reported attending ethics workshops at conferences enabled them to gain feedback about potential UCs from others (P1, P4, P17). For example, P9 specified that meeting with others at conference workshops allowed \emph{``leading researchers and also students [to] come together, [and] would help the community a lot in terms of figuring out [broader impacts].''} Additionally, two researchers mentioned their participation in university-level ethics board (P12, P18). In particular, P12 shared how their experience with volunteering at their university's ethics board enabled them to \emph{``identify problem[s] that a lot of people face...and provide a resource [of] people who had different research experience in diverse areas''} } P1 also expressed their enthusiasm for greater diversity in computer science labs: 
 
 \begin{quote}
 \emph{``I would [want] to have more people like even just people in my lab that have more diverse experiences and backgrounds. I think that we already do have that and that’s why I’m saying that I’ve noticed that this is super helpful when talking about unintended consequences.''
     }
 \end{quote}
 
\rr{Further, participants explained how several barriers (related to current academic practices, Section 5.3) prevented them from working with others. For example, P3 shared how they usually had enough time to ask for external help only to conduct ``sanity checks'' but otherwise lacked time to officially ``bring in outside experts.'' Other participants intentionally chose to collaborate within their own labs because it was assumed that their peers would offer sufficient feedback and suggest  potential UCs during lab meetings, P13 noted that their lab did not seek external collaborations. }

Participants also specified that diversity and inclusivity had directly impacted their research projects. In particular, P15, an accessibility researcher, shared how a recent collaboration \emph{``with a researcher with a disability''} helped their lab \emph{``[learn] a lot from their experience and [brought] in perspective from those communities.''} Additionally, P14 shared that lacking a diverse team of researchers on a past research experience led to a UC that users found \emph{``deeply offensive''}; P14 also highlighted how this experience led to future collaborations with more diverse teams and the use of participatory design methods. Generally, researchers agreed that diverse researchers could better help anticipate UCs by considering a greater variety of perspectives.

\subsection{Academic Practices Strain Efforts to Anticipate UCs}
\label{5.3}

Despite growing attempts to encourage anticipating UCs in research, we found that \textbf{the ``move-fast'' academic practice strains efforts to consider UCs}. Under this theme, we identified a list of barriers to elaborate how existing academic practices influence researchers’ actions and attitudes towards UCs. In particular, we describe how academic pressures to publish frequently impacts researchers’ considerations of UCs.

First, several participants described that they did not have enough time to devote to thinking about UCs. Participants felt that \textbf{considering UCs was additional work that might conflict with their goals to publish quickly}. To fully consider the social impacts, participants need to balance between moving fast and slowing down to brainstorm the future. Usually researchers chose to move fast, despite their hope to slow down ethically. P3 described their attitudes towards considering UCs while working on research projects: \emph{``
I think we all have a tendency to try to move fast and [think] `Oh no, I don’t want to stop and think about [unintended consequences], I want to keep going [...] to get the results.'''}
For example, P1 described that they prioritized the \emph{``technical or theoretical hurdles''} and considering UCs is \emph{``never one of those hurdles [...] never something that you explicitly approached with.''} 

Participants frequently commented that this academic pressure created incentives for producing more publications, rather than promoting ethical considerations. To some participants like P8, the result of this incentives system is the widespread belief that UCs are non-essential and \emph{``just adds more burden on the researcher.''} P19 explained: \emph{``In academia, the incentives are lacking [and] misaligned, which kind of feeds into that broader system and whole obsession over count metrics, so people try as hard as possible to publish as much as fast as possible.''}
Similarly, P2 commented: \emph{``
Incentives within [academic] systems may lead [researchers] to prioritize or de-prioritize things based on what’s most valuable: so in academia that will be publications.''}
P14, an associate professor in HCI, also scrutinized the research practices of AI and ML researchers, who \emph{``can start a project at the first one month and have a written paper in three months.''} They continued that this oftentimes leads to \emph{``unethical work that have high citation rates and high rates of turning out publications.''}

\textbf{The ``move-fast'' academic practice makes junior researchers difficult to consider UCs.} P7, a full professor at an R1 university, elaborated:
 
 \begin{quote}
     \emph{``[J]unior faculty, sometimes they don't have the time, the mind space, or the clarity to think about these issues, because [they’re] just busy doing other things. I don't fault them for that. It's just [I] don't think we're creating an environment such that they have enough oxygen to think about these issues. So I think that's something for us as a community to think about is just how do we, how do we build a space? And how do we build the time such that people are educated about these issues, and then have the opportunity to develop nuanced thoughts on them?''}
\end{quote}

 Ultimately, these comments by our \participants revealed that the unfortunate by-product of the academic pressure to publish inadvertently encouraged a system where researchers may fail to have the time for, or even see the value of, considering UCs in their research. 
 
 On a positive note, while \participants expressed frustration with the academic system, some also mentioned how things might be slowly starting to change. P15 shared an experience where they worked with a student who pointed out significant ethical considerations that they had overlooked for a research project using information learned from \emph{``a particular [ethics] class that that [the student] took''}.

 As P6, a full professor in NLP, told us:
 
\begin{quote}
\emph{``I have more hope for the new generation, the new generation of students, because they tend to be more concerned about it in general, which might be to do with the fact that they somehow learned about it during their college education, for example, they may have had a class about ethics and inequality.''} 
\end{quote}

Hence, with an increase in ethics courses, a general increase in awareness, and with systemic changes that mitigate barriers for considering UCs, academics may become more conscientious and incentivized to think about societal impacts of their work in the future. 

\section{Discussion}

Our findings demonstrate that computer science researchers in our study do not formally consider any potential societal impacts of their research innovations, despite perceiving it as important. 
Participants contemplated potential societal impacts of their innovations only in hindsight, such as when they were required to write a broader impact statement for a conference. They responded to individual incidents, such as when receiving participant feedback or getting bad press, instead of actively considering the topic. We contend that this observation is troublesome, as argued by Pillai et al.~\cite[p.2]{g2021co}: ``Unless ethics is integrated in every aspect of the design process and educational curriculum, it is bound to be an afterthought and thus inadequate in identifying and addressing ethical issues.'' 

While our participants generally suggested the need to proactively consider UCs, we identified various knowledge and structural barriers that currently prevent them from doing so.
First, participants felt that the lack of considering UCs is due to academic practices promoting fast progress and publications. In their eyes, the pressure to publish encourages researchers to de-prioritize and resist any form of stepping back and considering long-term effects. Several interviewees felt that spending time evaluating potential UCs slowed their momentum when conducting actual research. 

Second, we identified that researchers lack guidelines, tools,  and methods for considering UCs. None of our  participants reported using any of the existing tools for brainstorming about potential societal impacts, such as the Envisioning Cards~\cite{EnvisioningCards} or the Tarot Cards of Tech~\cite{tarotcards}. They also suggested that there are no  approaches or processes for thinking through the societal implications of their research that they felt genuinely satisfied with. 

 \begin{table}[t] 
    \footnotesize
    \centering
    \caption{Five Causes of Unanticipated Consequences by Robert Merton ~\cite{merton1936unanticipated}}
    \label{table:cause_merton}
    \centering
    \begin{tabular}{p{0.95\columnwidth}}
        \toprule
        \textbf{Ignorance}: Lack of knowledge, experience, expertise, and prudent investigation of a problem. \\
        \textbf{Errors}: Incorrect reasoning, analysis techniques, and interpretation of a problem. \\
        \textbf{Imperious immediacy of interest}: Actor's paramount concern with the foreseen immediate consequences excludes the consideration of further or other consequences of the same act. \\
        \textbf{Basic Values}: No consideration of further consequences because of the felt necessity of certain actions enjoined by certain fundamental values. \\
        \textbf{Self-defeating prophecies}: Predictions are frequently not sustained precisely because the prediction has become a new element in the concrete situation. \\
        \bottomrule
    \end{tabular}
\end{table}

Third, participants mentioned the lack of demographic diversity in collaborators and other academics, which they felt could enrich different viewpoints and experiences. This is a known structural issue in computer science, an occupational group that is heavily skewed towards white males~\cite{echeverri2018unintended}. Although the U.S. student population is becoming more diverse, faculty members and general academics remain predominantly white and male~\cite{bourabain2021everyday}. Furthermore, in a number of fields, most research published at high-profile venues is conducted in Western countries, with only Western participants~\cite{Linxen2021}. This means that research and innovation processes are distanced from the lives and experiences of many of their future users. Additionally, innovations can affect different kinds of people in unpredictable ways. This lack of diverse viewpoints and experiences characterized innovations that are biased against minority populations~\cite{pereidadiversity,hankerson2016does}. 
 
 Our exploration of the barriers to researchers' (in)actions resonates with Merton's five causes of UCs (see Table~\ref{table:cause_merton})~\cite{merton1936unanticipated} and enables us to place his theory in the academic context today. For example, Merton described how ignorance (i.e., a lack of knowledge and experience) can lead to UCs. We saw ignorance being part of the issue in the results of our study, with many researchers lacking the know-how to anticipate UCs and the experience of thinking about potential impacts from diverse perspectives. Merton also suggested that UCs can be caused by an ``imperious immediacy of interest,'' with people being driven to focus on the foreseen, desired consequences. As our participants mentioned, academic pressures to publish could exacerbate the desire to meet \emph{intended} consequences, such as developing a technology for the purpose of publication and/or getting a degree or promotion. They may therefore unconsciously or consciously ignore unintended effects. 


 Our results indicate that researchers would more routinely consider UCs if they felt that their research outcomes had a future impact on society. As it is, many deflect responsibility to those whose research they believe is more likely to influence products. The insight surprised us given that all interviewees were selected based on their work in applied research areas, and all had previously worked for research spin-offs, open-source projects, or companies. It suggests that there may be a gap between the impact academics think their work will have and the actual risk of UCs. However, the finding is in line with recent work~\cite{nanayakkara2021unpacking, Ashurst2022AIES} analyzing the text included in the broader impact statements for NeurIPS conference papers. Their results show that authors rarely state who is responsible for preventing negative impacts, and if they do, it is often a call for action rather than a statement of adopting personal responsibility.

 Several participants mentioned that they relied on their IRBs to alert them to potential UCs, wrongfully assuming that this ``falls under the same jurisdiction.'' While providing such structure or access could certainly help, it may risk replacing researchers' responsibilities with ``legalistic bureaucracy,'' a known criticism of the IRB~\cite{brown2016five}. This is the case for ethics checklists as well, such as those designed to guide practitioners' development of AI systems, which have been found to obfuscate responsibility while additionally being too abstract or ignored~\cite{madaio2020co}. Nevertheless, the finding emphasizes the need for computer science researchers to receive more guidance in thinking through societal implications, be it through an IRB-like structure or access to ethic experts who can provide guidance.

 The deflection of responsibility could be attributed to insufficient prior experiences with computing ethics and awareness of cautionary tales of innovations that had negative societal impacts. In fact, many participants cited their limited experience as a barrier to thinking about potential ethics pitfalls.  This may be slowly changing as more universities now offer ethics classes as part of their computer science curriculum~\cite{fiesler2020we}, which, as  several senior faculty interviewees noted, is how their students gain experience with considering these consequences.
 With an increase in ethics education and a generally heightened mindfulness of these issues due to media attention and public outcries, we may look back and be surprised that computer science researchers do not regularly consider ethical consequences in the research process. We remain hopeful, but our findings also show that overcoming barriers  will require a holistic approach to restructuring academia. In the following, we discuss how to better support researchers anticipating UCs. 

\section{Supporting Consideration of Potential UCs of Research Innovations}
 
 Our analysis suggests that there is still a long way to go before UCs are fully taken into account by researchers, despite recent calls to do so. Our work, however, does not contend that all computer science researchers should be ethicists to address the issue. Instead,
 by properly anticipating these incidents of varying severity~\cite{Greene2022AtomistOH}, researchers might also avoid suffering from their own reputation tarnishing and unexpectedly developing harmful technologies. 
 Releasing technology innovations into the wild without fully considering their effects on society is arguably a very large human-subject experiment with unknown outcomes. Our work reveals that considering societal impacts, at this point, is primarily an individual responsibility and that some form of oversight should be in place. 
 While structural changes may be needed to overcome these issues, we offer concrete suggestions for tangible next steps as follows.  \\
 
 \textbf{Collect and disseminate case studies of UCs that were the result of computer science research innovations.} Our findings suggest that participants sometimes felt their own research to be too prototypical and too removed from public release to cause UCs. Others believed that certain research, especially if the goal is to address a societal issue, was less likely to result in UCs. These findings suggest the need for information that could raise awareness of UCs that resulted from varied research innovations. Concretely, we suggest collecting and disseminating case studies of the societal impacts of computer science research, which can serve as an informational resource for teaching new and experienced researchers about prior work that has had (differential) negative effects on society. It is crucial that such case studies and other relevant resources demonstrate different aspects of UCs on innovations that might be otherwise overlooked. \rr{Moreover, collecting such resources should be an invitation to collectively learn from past mistakes instead of finger-pointing, and their submission should be encouraged, such as by providing specific awards.}  \\ 
 
 \textbf{Develop tools that support learning and brainstorming about UCs.} Our study shows that researchers need more structured guidance to think through potential societal impacts. While some tools already exist, such as the Tarot Cards of Tech or Envisioning Cards, these often target practitioners and are not always suitable for research artifacts. They also require much time and cognitive effort to brainstorm about UCs. Moreover, anticipating the downstream uses and effects of research innovations is a notoriously difficult problem~\cite{prunkl2021institutionalizing}. Since a majority of our participants relied on their peers' knowledge of prior UC experiences, creativity support tools, such as those that cluster ideas from previous users to support the generation of more diverse ideas~\cite{siangliulue2015toward}, could promote sharing of UCs within communities. Such collaborative ideation tools could also serve as one approach for engaging citizen scientists in the research process. 
 \\ 
 
 \textbf{Increase access to input from diverse people in research design and development.} Our results show the benefits of more demographic diversity among researchers, so their projects can be conceived and evolve with input from people with varying values and experiences. \rr{In particular, several participants mentioned how, especially when attending alongside a diverse panel of researchers, participating in ethics workshops at conferences supported conversations and considerations about ethics in their research.} This suggestion also reflects discussions by Kling ~\cite{kling1991computerization}, who encouraged computer science practitioners to apply diverse perspectives from other fields (e.g., the social sciences) to recognize potential societal implications, and Hankerson et al.~\cite{hankerson2016does}, who suggested that hiring more diverse people at every level could help mitigate racial biases in technology. 

While these suggestions are long-term goals, a more immediate suggestion, inspired by human-centered design practices, is to engage with diverse participants of varying backgrounds, skills, and characteristics to ideate and test new technologies. For example, researchers have found that unrepresentative participant sampling may lead to the creation of non-generalizable technologies that amplify existing inequities \cite{robertson2018, Linxen2021}. Similarly, to prevent further social inequity in technology, we also propose the use of participatory and co-design methods, which have been shown to more equitably and effectively inform technologies by directly identifying user needs~\cite{Walsh2018, Tojo2022}.

Second, we suggest inviting the public to routinely shape research ideas and ongoing projects with comments and feedback. For example, Johnson and Crivellaro designed a community panel to create dialogic spaces that foster critical engagements with technologies and social issues for the purpose of reviewing research proposals on HCI~\cite{johnsonopen}. Many citizen science projects witnessed radical improvements due to public contributions~\cite{hand2010citizen}. 
These projects offer a give and take, with researchers receiving help collecting, annotating, and analyzing data in exchange for an educational gain for citizen scientists~\cite{miller2020introducing, labinthewild}. A similar involvement of citizen scientists in evaluating the societal impacts of research innovations (at all stages of the research process) could benefit both researchers and the public. 

Of course, such involvement requires precautions. \rr{For example, participatory methods might bring unequal power dynamics and risk over-reliance on citizen scientists who may not be experienced and knowledgeable in examining these issues~\cite{Birhane2022PowerTT}.} Researchers may also be hesitant to openly discuss their ideas before they are published. This may be mitigated by allowing the preregistration of ideas (similar to what has been encouraged for research studies~\cite{cockburn2018hark, registration}). Overall, we believe that allowing the public to have a voice in evaluating the societal implications of research projects on them will be a challenging but rewarding endeavor. \\


\textbf{Increase incentives to investigate UCs at various stages of the research process.} 
Our findings suggest that academics may have a lower incentive to investigate UCs than companies do, because they often do not perceive their research as immediately affecting large numbers of users and because they are less likely to face direct public backlash. While any additional task will be perceived as burdensome, we can learn from several efforts to improve scientific practices. For example, conferences have included tracks that auto-accept papers if researchers preregister their studies~\cite{icsme}. Several venues have also encouraged the replication of studies to improve the quality of science~\cite{wilson2011replichi}. Similarly, computer science venues could encourage the publication of auditing research artifacts~\cite{raji2020closing} by providing special tracks. To support such early audits, these venues could additionally allow for submissions of experience reports that describe how a research project was modified or even canceled due to the discovery of unanticipated societal impacts. According to our participants, researchers would greatly benefit from learning about such experiences to prevent UCs in the future. 

Additionally, conferences, research methods classes, and advisors should reframe finding negative UCs as an opportunity. Oftentimes, analyzing a project for disparate outcomes can ``serve as a starting point for another research project''~\cite{IUI2022}. For example, prior work on language models has suggested that discussing the potential harms of this technology can ``stimulate efforts to study and mitigate them''~\cite[p. 34]{brown2020language}. Similarly, research communities should consider rewarding researchers for uncovering and addressing societal consequences and provide venues for openly debating whether research efforts should continue, be abandoned, or change direction.

Finally, as our participants suggested, adding ``lessons learned'' to existing research publications should be made possible in digital libraries to allow others to learn from these experiences. Collecting and sharing these experiences with researchers may also help overcome the common ``not me'' attitude that we have seen in our study by showing researchers that no matter how well intentioned, almost all research can impose negative consequences that must be considered. \\

\textbf{Front-load considerations of societal impacts.} 
 Recently, conferences require or nudge authors to consider UCs, e.g., by including broader impact statements in papers. However, critics have raised concerns that these statements contain speculative fiction and are published too late in the research process~\cite{Adar2021}. 
 Front-loading considerations of societal impacts in the research process can avoid the point of no return, as some of our participants described it. One possibility is encouraging pilot studies to confirm ethical practices before conducting large-scale human-subjects experiments~\cite{Truong2017}. Another possibility is to require submitting an analysis of potential UCs to funding agencies along with research proposals. Funding agencies commonly request broader impacts statements, but these are mostly used to advertise the positive impacts of a research project rather than to reflect on potential negative consequences.
 
 Two challenges must be addressed to make this happen: First, authors of a grant proposal would need adequate guidelines and tools to think through societal impacts. Second, funding agencies and their reviewers would need to learn how to evaluate such societal impact statements. A similar process was successfully tested at an academic institution~\cite{Bernstein2021EthicsAS}, though it was cautioned that it can be perceived as ''burden`` without additional scaffolding. 
 
 Therefore, one immediate next step is to invest in research efforts that can provide guidance to authors and reviewers. Institutions could further assist researchers with regular consultation by hiring an ethics advisor similar to the ethics consultation service in most medical programs~\cite {bioethics, medical}.  With the onset of more advanced technology applications, having a technology ethics advisor on staff to answer researcher questions and review proposals and papers might alleviate burdens for researchers on their own.\\

\textbf{Provide guidelines for reacting to societal impacts.}
 In addition to the lack of guidelines for \emph{considering} UCs, our participants frequently mentioned the lack of guidelines for \emph{reacting} to them once they were discovered. Their hesitation to consider societal impacts may come from knowledge barriers regarding anticipating and reacting to UCs. In line with Knight's recommendation to addressing UCs, we offer two suggestions that aim to ``increase knowledge'' about UCs and combine uncertainties within a large-scale community ~\cite{knight1921risk}. First, we advocate creating opportunities where researchers with similar research topics can share their experiences with considering and addressing UCs. These opportunities, which could be digital platforms (e.g., online forums) or physical venues (e.g., conferences and workshops), would not only support researchers by recognizing common UCs within their sub-disciplines but would also create community and dialogue around UCs. Second, funding agencies could provide resources similar to the IUI 2022 program chairs, who have put together a list of links to papers and tools for authors to audit their projects~\cite{IUI2022}. Providing them with checklists and tools to evaluate these analysis reports will be essential to avoid ethics washing~\cite{prunkl2021institutionalizing}. Altogether, providing guidelines for such decisions is desirable; we believe that HCI researchers, given the field's interdisciplinary nature and focus on societal impact, are uniquely equipped to drive research in this direction. 
\section{Limitations and Future Work}
 Our work provides insight into the barriers that prevent computer science researchers from considering and addressing UCs of their work. One limitation of our work is the relatively limited sample size and diversity of participants in this study. We selectively spoke to North American computer science researchers from R1 research universities. 
 \rr{Because academia can be structured differently across countries, our findings may not reflect either the resources or challenges that academics in other parts of the world encounter when considering societal impacts of their work. The focus on R1 institutions additionally means that we cannot conclude the habits and challenges in considering UCs are similar to R2 or special focus institutions.}  Moreover, because we employed purposive sampling, our findings may not generalize to all computer science researchers; in particular, we suspect there may be differences if researchers come from marginalized or minority backgrounds, if their research addresses ethics, or if their research is further removed from applications. Future work is needed to investigate whether our findings and suggestions for supporting computer science researchers in anticipating UCs generalize to other research institutions within the US and to other countries.

 An additional limitation is our focus on academic computer science researchers whose applied research products have led to systems used by the general public. \rr{This might have overlooked opinions by researchers in industry who may be subject to different policies and organizational structures within their companies.} We might have missed opinions by researchers in other fields that are deeply affected by UCs in technology. Future work should explore how computer science researchers in their specialized sub-disciplines with varying demographic backgrounds, work experiences, and research experiences enrich our findings.

 Our results are also impacted by the possibility of response bias, a common issue in interview studies. Given the heightened awareness of societal implications and the blame associated with it, the risk is that participants may have appeared more concerned about UCs than they actually are and may have downplayed any UCs that they have experienced themselves. We did not perceive this as an issue in their responses, but our findings on the perceived importance of proactively considering UCs should still be taken with a grain of salt. Future work could build on these findings with an anonymous survey or with longitudinal studies on people's motivations and actions for considering UCs.
 
 Additionally, our work mainly focuses on how researchers anticipate unintended consequences. Future work should explore the opportunities and challenges that researchers encounter when reacting to unintended consequences in greater detail. Based on our initial findings from this work, we assume that researchers face similar structural and knowledge barriers when responding to unintended consequences during the research process.
\section{Conclusion}

In this paper, we share how computer science researchers anticipate the UCs of their research innovations and the challenges they may face from doing so throughout the research process. Through our interviews with 20 computer science researchers affiliated with North American research universities, we learned that our interviewees do not have a specific process or strategy for considering UCs. 
Our interviews surfaced two major barriers to anticipating UCs for researchers including a lack of formal methods and guidelines for anticipating them and academic practices that promote fast progress and frequent publications.
Based on our findings, we outline key opportunities to support researchers in these efforts by incentivizing the investigation of UCs throughout the research process, the creation of new tools to support brainstorming, and the implementation of reactionary guidelines. We hope that our work will encourage further discussions on how academic innovators can be supported in the prevention of unintended effects of technology on society. 

\begin{acks}
We thank our participants and the anonymous reviewers for their valuable feedback and suggestions. We also thank Sandy Kaplan and Mary Peng for their help revising this paper. This work was funded by the National Science Foundation under award IIS-2006104. Any opinions, findings, conclusions, or recommendations expressed in our work are those of the authors and do not necessarily reflect those of the supporter.
\end{acks}

\bibliographystyle{ACM-Reference-Format}
\bibliography{bibliography}

\appendix

\end{document}